# Core determinants of quality criteria for mhealth for hypertension: evidence from machine learning instruments

Danielly de Paula*, Ariane Sasso, Justus Coester, Erwin Boettinger

Hasso Plattner Institute, Potsdam, Germany

**Abstract**

Uncontrolled hypertension is a global problem that needs to be addressed. Despite the many mHealth solutions in the market, the nonadherence relative to intended use jeopardizes treatment success. Although investigating user experience is one of the most important mechanisms for understanding mHealth discontinuance, surprisingly, the core determinants of overall user experience (i.e., positive and negative) about mHealth apps for hypertension are unknown. To address the mentioned gap in knowledge, this study adopts the computational grounded theory methodological framework and employs advanced deep learning algorithms to predict core quality criteria that affect overall user experience of hypertension apps published in the Apple App Store. This study contributes to theory and practice of designing evidence-based interventions for hypertension in the form of propositions and provide valuable managerial implications and recommendations for manufacturers.

## 1 Introduction

Despite current technological and medical advances, the world is experiencing an increasing burden of noncommunicable diseases – also known as chronic diseases - which counts for 74% of all deaths globally (WHO). Hypertension – or elevated blood pressure - is one of the most common chronic diseases and it is largely associated with unhealthy lifestyle such as low physical activity, smoking, and unhealthy diet (Cao et al. 2022). Globally, the number of adults aged 30–79 who are hypertensive has doubled which leads to a burden on the healthcare system that is already overstretched (Zhou et al. 2021). Once diagnosed, patients need to follow a treatment protocol while changing their lifestyle. Accordingly, the responsibility for day-to-day disease management gradually shifts from health care professionals to the individual (Alessa et al. 2018). As the healthcare system is transitioning from reactive care to a more proactive care (Ghose et al. 2022), investigating how to design effective self-management solutions for hypertensive patients is a pressing global challenge (Angell, De Cock & Frieden 2015; Choi et al. 2020; Zhou et al. 2021).

Mobile health (mHealth) apps, wearable sensors, and devices for home blood pressure monitoring offer great potential in the self-management of hypertension through the involvement of individuals in the self-tracking of behavioral and biological information (Hui et al. 2019). Despite tremendous promises and expectations, little has been achieved so far regarding how mHealth apps can indeed improve health and behavior outcome of hypertensive patients (K. Liu, Xie & Or 2020).

---

*Corresponding author: danielly.depaula@hpi.de

To date, scholars have mostly focused on using mHealth for hypertension that addresses medicament monitoring neglecting fundamental variables for effective self-management such as behavioral (e.g., diet, physical activity), knowledge (e.g., empowerment over the condition), and psychological outcomes (e.g., distress) (K. Liu et al. 2020). To develop the next generation of mHealth for hypertension that positions the user in a more active role in their own care, it is important to identify and prioritize required product improvements.

Although the topic of self-management of hypertension is not novel (Alessa et al. 2018; Hui et al. 2019; Kazuomi, Harada & Okura 2022), scholars are concerned that the majority of the studies report that users do not adhere to the apps as intended (Choi et al. 2020; Jakob et al. 2022). Adherence is defined as "the degree to which the user followed the program as it was designed" (Donkin et al. 2011:2). The nonadherence relative to intended use jeopardizes treatment success which might explain why World Health Organization recently reported that hypertension is poorly controlled worldwide (World Health Organization (WHO) 2021). While some studies report that the presence of personalized goal-setting were found to be favorable in reducing blood pressure, whereas body monitoring features yielded limited efficacy (K. Liu et al. 2020), the scientific body of literature lacks a concise conceptualization of what are the core determinants of the overall user experience (Hui et al. 2019; K. Liu et al. 2020; Kazuomi et al. 2022) .

While the *what* and the *why* underlying user experience is often understood through rigorous social science approaches such as interviews and surveys (Ojo & Rizun 2021), we argue that analyzing online reviews – also known as user-generated content (UGC) - through computational methods enable us to investigate *how often* a feedback occurs. Understanding the frequency that a topic is mentioned allow us to identify core determinants of user experience. Accordingly, this study aims to analyze user-generated content in the current landscape of hypertension apps available in the Apple App Store to determine latent topics and predict which of those topics are the core determinants of quality criteria, and consequently user experience. To achieve our goal, we draw on the quality criteria from the mHealth evaluation framework proposed by Hensher et al. (2021) as we argue that the framework provide a blueprint onto which topics can be mapped, helping to identify core themes, aiding in prioritization (high actionability). The following research questions (RQ) guided our study: what topics presented in online reviews from an app store are related to quality criteria from the health app evaluation framework (RQ1), and which are the core determinants of quality criteria (RQ2)?

As part of the methodological underpinnings of this work, we employed Computational Ground Theory (CGT) which is as a comprehensive methodology for generating novel design knowledge from unstructured textual data (Berente & Seidel 2014; Nelson 2020). The methodological framework combines human skills in interpretation with computational techniques to analyze large unstructured data. Studies using the framework are scarce, and to the best of our knowledge, our study is among the first efforts to operationalize CGT to generate insights from online reviews from mHealth applications published in applications distribution platforms. In particular, first, we developed an open-source tool to download relevant user reviews of mHealth apps for hypertension from the Apple App Store. Then we used a deep learning architecture - BERTopic - with a traditional machine learning method — Random Forest Classifier —to identify the core determinants of quality criteria based on the model proposed by Hensher et al. (2021). Our findings indicate that promoting an effective communication between users and health care professionals is associated with high user-rating and therefore positive user experience, whereas the lack of interoperability and privacy concerns is associated with low user-rating and therefore negative user experience. Moreover, current solutions have focused mostly on self-monitoring which is only one aspect of self-management. More focus on providing just-in-time interventions that are aligned with behavioral theories is needed. This study contributes to the body of knowledge of designing evidence-based interventions for hypertension apps by predicting core quality criteria that influence the overall user experience. Additionally, it contributes to practice by offering recommendations for product developers through a set of features that the users are most concerned with. Methodologically, it (a) demonstrates how to operationalize CGT in the context of user reviews from the Apple App Store, and b) offers an open-source tool and machine learning model that can be used to extract and prioritize knowledge about other health conditions. The paper is organized as follows. The next section explores the extant literature on self-management of hypertension in the digital era and current methods for data collection and data analysis to extract knowledge from user reviews. Then, the section is followed by the description of the methodology employed in our work. We then present our results and discuss how our findings contribute to both theoretical and practical knowledge of factors that matter when improving the overall quality of mHealth for hypertension.

## 2 Literature Background

### 2.1 Self-Management of Hypertension in the Digital Era

The involvement of patients in the management of their health care is referred to as self-management (SM) which is defined as the "individual's ability to manage the symptoms, treatment, physical and psychosocial consequences and life style changes inherent in living with a chronic condition" (Barlow et al. 2002:178). In the context of chronic diseases, such as Hypertension, evidence suggests that involving the patient in the management of their own care is fundamental because it not only enables them to make informed decisions but also enhances their self-efficacy (i.e., capacity to undertake specific health behaviors) (Savoli, Barki & Pare 2020).

Hypertension – a highly prevalent disease affecting one in three adults – and the number one risk factor for heart disease and stroke, both of which are leading causes of death (Muntner et al. 2020). Hypertension management requires lifelong self-care of patients by health care professionals for medication dosing (Márquez Contreras et al. 2019), and in many cases, daily support for behavioral changes are needed – however; unlikely to come from health care professionals (Alessa et al. 2018). mHealth is increasingly being used to help address these challenges through self-management features such as information exchange, health literacy and peer support without the constrains of time and geography (K. Liu et al. 2020). Although there is an increase in affordable home blood pressure monitors and there are several mobile apps that support patients with hypertension, blood pressure remains to be poorly controlled (World Health Organization (WHO) 2021). That happens because user acceptance of hypertension apps decrease over time as a reflection of the users' app-engagement experiences, app technical functionality, and/or design features leading to significant drop outs (Choi et al. 2020; Jakob et al. 2022). Considering that behavioral change for hypertension control requires a long-term commitment, the nonadherence relative to intended use is critical and highlights the necessity of developing more effective models, best practices, and interventions for hypertension apps. In particular, scholars mention that in order to prevent significant drop outs (Horneber & Laumer 2022) and increase the effectiveness of mHealth apps it is necessary to understand the factors that act as barriers to or facilitators of positive user experience (Jakob et al. 2022). Although understanding user experience is one of the most important mechanisms for investigating mHealth discontinuance, surprisingly, the core determinants of user experience (i.e., positive and negative) about mHealth apps for hypertension are unknown (K. Liu et al. 2020; Jakob et al. 2022).

While most feedback concerning user experience is collected either through psychometrically sound surveys (Ojo & Rizun 2021), or rigorous qualitative approaches such as interviews and focus groups (de Paula, Zarske Bueno & Viljoen 2021), increasing attention has now been paid to investigating online reviews – i.e., user-generated content (Lukyanenko et al. 2019). In this study, we argue that harnessing user-generated content from application distribution platforms (e.g., Apple App Store) enable us to determine the core quality factors that affect user experience. To investigate core quality factors that determine user experience through user reviews about hypertension apps, we rely on computational techniques. In the next section, we analyze studies that have employed computational techniques to extract knowledge from user-generated content.

### 2.2 Extracting Knowledge from User-Generated Content About mHealth

User-generated content is digital information produced in a variety of forms—such as tweets, product reviews, forum posts – by members of the public who are often unpaid and not affiliated with the organization (Goes, Lin & Yeung 2014; Lukyanenko et al. 2016; Lukyanenko et al. 2019). The digital information reflects user feedback that organizations can then utilize to better understand their customers, competitors, products, and services. In particular, scholars suggest that user feedback through online reviews provide important demand-side knowledge for manufacturers to improve product quality (Zhou et al. 2018), which enables the users to move from "passive listeners" to "active players". Due to its potential for organizational decision-making, the topic has attracted interest from many disciplines such as psychology (Ding et al. 2021), medicine (Alsheref 2019), and information systems (IS) (Zhou et al. 2018). Although UGC in online reviews often lack information on basic socio-demographic characteristics of users and are often large, complex, messy, and unstructured, when successfully harnessed, these reviews have the potential to provide relevant information for healthcare manufacturers to improve product quality (Ruelens 2021).

Recent developments in computational approaches are promising and are enabling researchers to uncover many interesting ways to extract knowledge from UGC. In particular, topic modelling is a probabilistic text mining technique based on an unsupervised machine learning approach that enables inductive and automated discovery of topics in large amounts of texts (Debortoli et al. 2016). The latest studies in topic modelling are applied to many contexts, such as understanding collaboration in asynchronous online discussions (Eryilmaz & Thoms B Ahmed 2022), identifying determinants of service quality from a hospital review platform ((Ojo & Rizun 2021), and investigating free texts in a professional women online network ((Schmitt et al. 2020). However, to the best of our knowledge, within IS research, there is no study that uses computational techniques to analyze user-generated content with a focus on the mHealth market. As a consequence, we do not know what techniques for data collection and analysis are reliable to be used in the context of online reviews in application distribution platforms. For instance, Latent Dirichlet Allocation (LDA) has been shown to be reliable for long texts, however; it is unclear whether it is recommended to be used for short texts. Moreover, topic modeling alone does not explain text data, it is necessary to draw upon a lexicon framework to extract knowledge.

When using topic modelling for analyzing online reviews about mHealth, we argue to use the model of Hensher et al. (2021) as a lexicon framework through which we can engage in sensemaking of the unstructured user reviews. Hensher et al. (2021) identified ten common quality criteria that are important to promote sustained usage of mHealth, these are: (1) Clarity of purpose of the app, (2) Developer credibility, (3) Content/information validity, (4) User experience, (5) User-engagement/adherence and social support, (6) Interoperability, (7) value, (8) Technical features and support, (9) Privacy/ethics/legal, and (10) Accessibility. The dimensions were created based on several established mHealth evaluation frameworks such as MARS and uMARS (Stoyanov et al. 2016) and provide guidance for scholars to analyze user experience. Additionally, we propose to combine the results of the topic modelling with Random Forrest Classifier (Breiman 2001) in order to predict which of the topics are the core determinants of quality criteria for hypertension apps. In the next section, we explain our methodological choices to identify and prioritize knowledge in user reviews.

## 3 Research Design

In this study, we applied Computational Grounded Theory which provides guidance for deriving novel design knowledge from large amount of data. An abstract approach of CGT has been proposed in IS research (Berente & Seidel 2014), later applied as a three-step approach (Nelson 2020), and most recently operationalized in details for the context of patient experience with hospitals (Ojo & Rizun 2021). For this study, while we followed the recommendations from IS scholars on how to operationalize CGT (Berente & Seidel 2014; Ojo & Rizun 2021), we also proposed new techniques for data collection and analysis to be used in the context for analyzing unstructured user reviews from the Apple App Store. For instance, while previous studies (Nelson 2020; Ojo & Rizun 2021) applied a three-step approach, our research design comprises four steps, which are explained as follow.

Considering that CGT had not been applied to analyze texts from app reviews before, in *Step 1 - Tool Development,* we developed a tool namely Apple Store Analytics to enable us to determine search criteria to select relevant apps and download relevant reviews. Our tool served as the main data collection technique for this study. Then, in *Step 2 - Pattern Identification*, we used BERTopic which is topic modelling technique that leverages BERT embedding to create clusters of latent topics based on a large dataset. Although IS researchers have mainly applied LDA, BERTopic is known for having a better performance for short texts than LDA (Alhaj et al. 2022). In *Step3 - Lexicon Filtering*, we mapped the latent topics to the quality criteria from the app evaluation framework (Hensher et al. 2021). Finally, in *Step 4 – Predicting Core Quality Criteria*, we analyzed which quality criteria is the strongest factor that influences the user's quality rating in an effort to understand the core determinants of overall user experience. Figure 1 illustrates our research design.

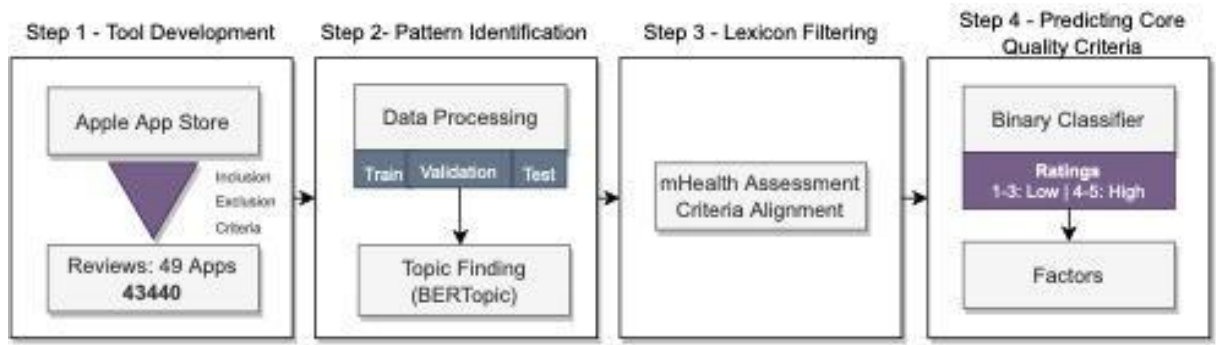

*Figure 1. Research Design: From Tool Development to Theory Building.*

## 3.1   Data Collection

To enable us to collect the reviews from the Apple App Store, the first step of our research design was to develop the tool Apple Store Analytics. The Apple Store Analytics uses the iTunes Search API[1] to search for relevant apps. Based on search criteria, the tool exports in JSON and XLSX formats a list of apps which contains relevant information such as the app ID, the app name, supported name, app genre, app page, company website, price, and description. Having set a list of app ids, through the iTunes RSS Feed we collect reviews. Once the reviews are collected, the tool detects the language of the reviews using fastText, and then automatically translates all non-English reviews to English using Google Translate. The tool can also be configured for other languages. At the end, a JSON file is generated which can be used for further analysis. When downloading the reviews, the authors' name from the reviews can be removed. The development environment used for this study was Jupyter Notebook. It is important to note that the tool has some limitations due to the limitations of the iTunes Search API. For instance, there is a maximum of 200 apps searchable per search per country (in our study, the average was 147 apps). Moreover, a maximum of 500 reviews are retrievable per country per app (10 pages of 50 reviews each) and the iTunes API now limits queries to about 20 calls per minute. Unlimited API, to date, is not publicly available. Access to the Apple Unlimited API namely Enterprise Partner Feed (EPF)[2] requires login credentials and an active partnership with the partners program. Finally, our newly developed Apple Store Analytics tool is open source and available for download on GitHub [3].

After the development of the tool, we defined search criteria to help us screen apps and identify those that are relevant to answer our RQs. First, we ran searched_apps script accessing the iTunes Search API using the search term "blood pressure" OR "hypertension" (#1). The search looks for the keywords in the description, name, and release notes of the apps on a global level. Moreover, all 175 iTunes countries (#2) and all languages were included (#3). As mentioned previously, the maximum search limit was 200 apps per search country. The processing time took 14 minutes and returned 626 apps. As exclusion criteria, we excluded apps that have less than 100 ratings (#5) due to the very few reviews available. Additionally, we excluded apps with GenreID: 6014 (Games) (#6). As a result, we obtained a list of 196 unique apps.

---

[1] https://developer.apple.com/library/archive/documentation/AudioVideo/Conceptual/iTuneSearchAPI/index.html

[2] https://affiliate.itunes.apple.com/resources/documentation/itunes-enterprise-partner-feed/

[3] https://github.com/hpi-dhc/apple-store-analytics

Furthermore, we decided to only include apps that offer connection with wearables (#4) (e.g. Apple Watch, blood pressure monitor). To address that, an additional set of keywords were defined which includes `Apple Health', `Bluetooth', `Connect', `Device', `Record', `Sync', `Watch', `WiFi', `iHealth'. Moreover, we added the names of the most popular blood pressure monitoring devices in the market (e.g., Withings and Omron). Considering that having additional keywords is not sufficient to achieve our criteria #4, we decided to manually analyze all pre-selected apps through their App page or Company website which was already listed in the search result. As a result, we obtained 49 apps that either have blood pressure monitoring as their primary function or act as a secondary feature in apps that monitor other conditions, such as diabetes. Then, we ran process-comments script to download the reviews, remove the author's name and to translate non-English reviews and titles to English. Our final dataset contains 43440 processed reviews from 49 applications and the average review length (summing the body and the title length) is 36.49 words. In total, 50 languages were detected. 78% of the reviews were written in five languages: English (21630), German (4486), French (2613), Spanish (2563) and Chinese (2056). 75% of the comments are from two continents: Europe and North America. Table 1 illustrates our inclusion and exclusion criteria.

| # | **Inclusion Criteria** |
|---|---|
| 1 | Search terms were "blood pressure" OR "hypertension" |
| 2 | All iTunes country stores |
| 3 | All languages |
| 4 | Connection with wearables (e.g. Apple Watch, blood pressure monitor |
| # | **Exclusion criteria** |
| 5 | Less than 100 global ratings |
| 6 | GenreID: 6014 (Games) |

*Table 1. Inclusion and Exclusion Criteria for App Search.*

## 3.2 Data Analysis

Traditionally, Latent Dirichlet Allocation (LDA*)* is used as a topic modelling approach to derive topics and insights from large unstructured text. In this work, however, we used a technique based on BERT to leverage the capabilities of pre-trained models and better represent contextual information (Devlin et al. 2018). Following the success of (Alhaj et al. 2022) in analyzing short text and the robust baseline set by (Grootendorst 2022), we selected BERTopic[4] as the technique to analyze the reviews recovered from our Apple Store Analytics tool.

BERTopic uses three phases to produce a topic distribution based on a set of documents—user reviews in our case. First, it creates from each user review a numerical vector representation of the text (i. e. embeddings). Next, it reduces the dimensionality of the embedding vectors using the UMAP algorithm (McInnes, Healy & Melville 2018), and then it clusters the UMAP embeddings with HDBSCAN (Campello, Moulavi & Sander 2013) to find similar text that composes a topic. The ideal number of topics is initially inferred by BERTopic. In the last step, the importance of words within each topic is defined according to a variation of the Term Trequency-Inverse Document Frequency (TD-IDF) measure named c-TF-IDF (Grootendorst 2022). As seen in Equation 1, this measure treats the documents from each cluster as a single document class *c* in order to show the importance (or weight *W*) of term *t* per class instead of per document. Those classes are in the end our topics that can be merged later from the least common one to its closest similar topic.

---

[4] https://github.com/MaartenGr/BERTopic

$$W_{t,c} = tf_{t,c} * \log(1 + A/tf_t) \tag{1}$$

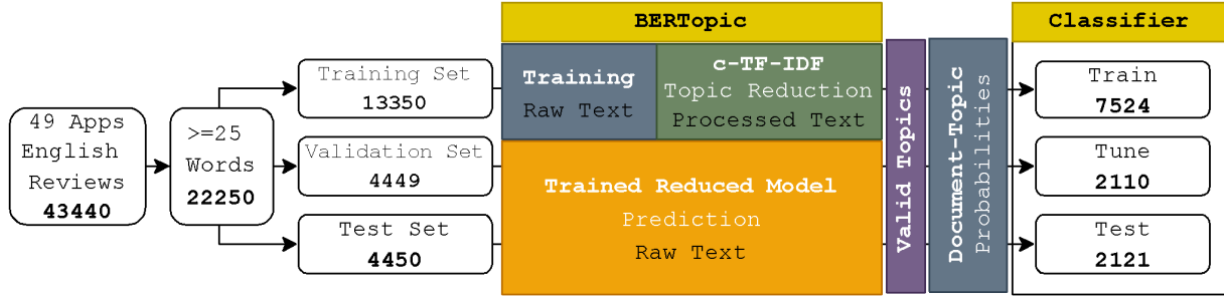

*Figure 2. Data Analysis Pipeline. After filtering, and splitting the dataset, the BERTopic model was trained. The document-topic probabilities were used as input to the classifier.*

Before feeding the dataset into BERTopic, we filtered all reviews that had less than 25 words as seen in Figure 2. We used the scikit-learn library (Pedregosa, Varoquaux & Gramfort 2011) to split the original dataset (43440 reviews) into training, validation and test sets (80%, 20% and 20%). One empty review was excluded from the validation set after the pre-processing step. We used the raw English text to train the BERTopic model and a pre-processed version of the text was applied to improve the results from the c-TF-IDF measure. We removed stop words using the Natural Language Toolkit (NLTK) library (e.g. 'the', 'and') (Bird, Klein & Loper 2009) together with a list of common words in the English language and terms such as 'app' and 'html'. Numbers were also removed from the pre-processed texts. Finally, we used lemmatization to reduce word variants to a common base form (e.g. 'accessibility' and 'accessible' to 'access') and selected only the resulting nouns.

After the training and reduction process, 30 topics were identified and were used in the following steps of our research design. To arrive to this number of topics we tried to maximize the topic coherence using the Normalized Pointwise Mutual Information (NPMI) metric (Bouma 2009) and topic diversity using the percentage of unique words (Dieng, Ruiz & Blei 2020) in a topic as seen in Figure 3. We chose the number of topics with the highest coherence score and still acceptable diversity. Moreover, 30 topics made more sense to the human reviewers than 40. This number of topics lies in the range of 10 to 50 topics, proposed as suitable for human interpretation (Debortoli et al. 2016; Schmitt et al. 2020).

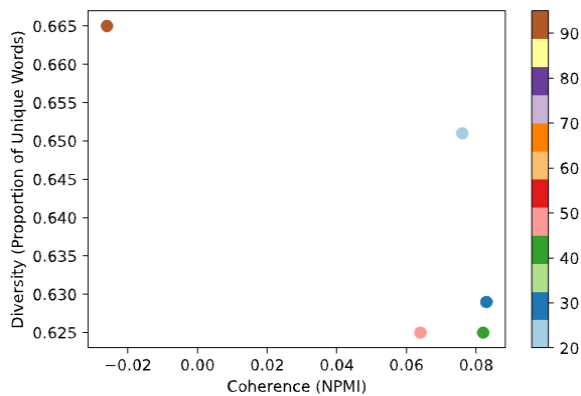

*Figure 3. This figure depicts the NPMI coherence score in the x axis and the topic diversity in the y axis. The original output of BERTopic was 95 topics which were later reduced to 30*

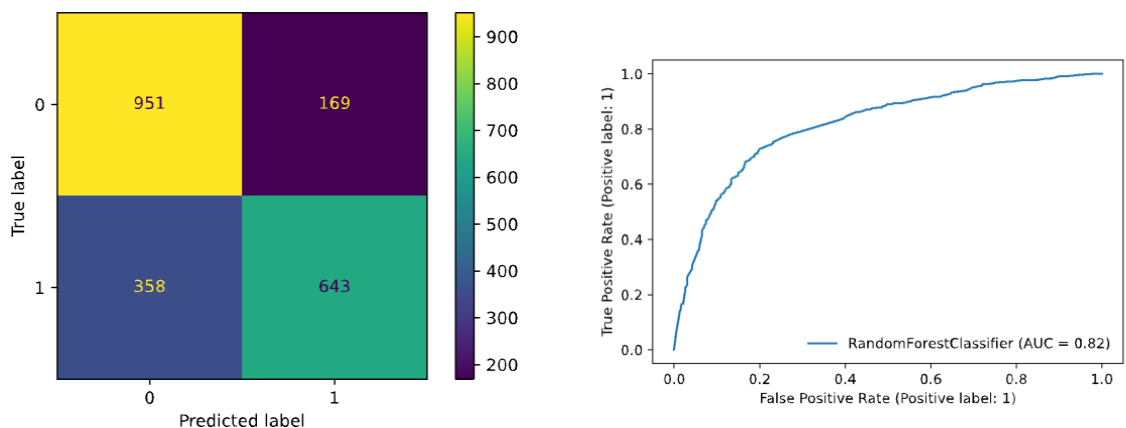

*Figure 4. Confusion Matrix and Area Under the ROC Curve*

The third step of our research design, lexicon filtering, comprises the interpretation and sense-making of the valid topics that the BERTopic revealed upon the training set. To perform this step, the following analysis was carried out: (1) the definition associated with each of the Mobile App Evaluation Framework (Hensher et al. 2021) were reviewed; (2) two scientists independently mapped each of the 30 identified topics to exactly one dimension of the framework and discussed until common agreement was achieved.

In the last step of our study, Predicting Core Quality Criteria, the Random Forests (RF) algorithm was applied to predict the influence of the topics regarding the quality rating provided by the users. The user ratings were classified into two classes - Low (rating values from 1 to 3) and High (rating value of either 4 or 5). For the classifier, the two classes were represented by either 0 (low rating) or 1 (high rating). To evaluate the performance of our classifier, we used a Confusion Matrix (see Figure 4) and to investigate the probability of each observation belonging to a class, the Area Under the ROC Curve (AUC) was computed. A Confusion Matrix is an error matrix where y-axis represents the True label (i.e., the instances in an actual class), and the x-axis represents the Predict label (i.e., instances in a predicted class). The RF algorithm is a commonly used tool for classification with high dimensional data, where the final prediction is then the majority vote (for classification) of the predictions of all trees in the forest (Breiman 2001). AUC measures the ability of the classifier to distinguish between classes where the higher the AUC the better the performance of the model. When AUC > 0.5, the classifier is able to detect more True positives and negatives than it would do by chance. As shown in Figure 4, the AUC for our classifier is 0.82 and can be considered good (Carter et al. 2016).

# 4 Findings

## 4.1 Mapping topics to the health app evaluation framework

Figure 5 presents a detailed overview of the 30 topics that we identified mapped with the dimensions of the app evaluation framework (Hensher et al. 2021). Below we present the most interesting findings from each dimension following a ranking of the number of reviews.

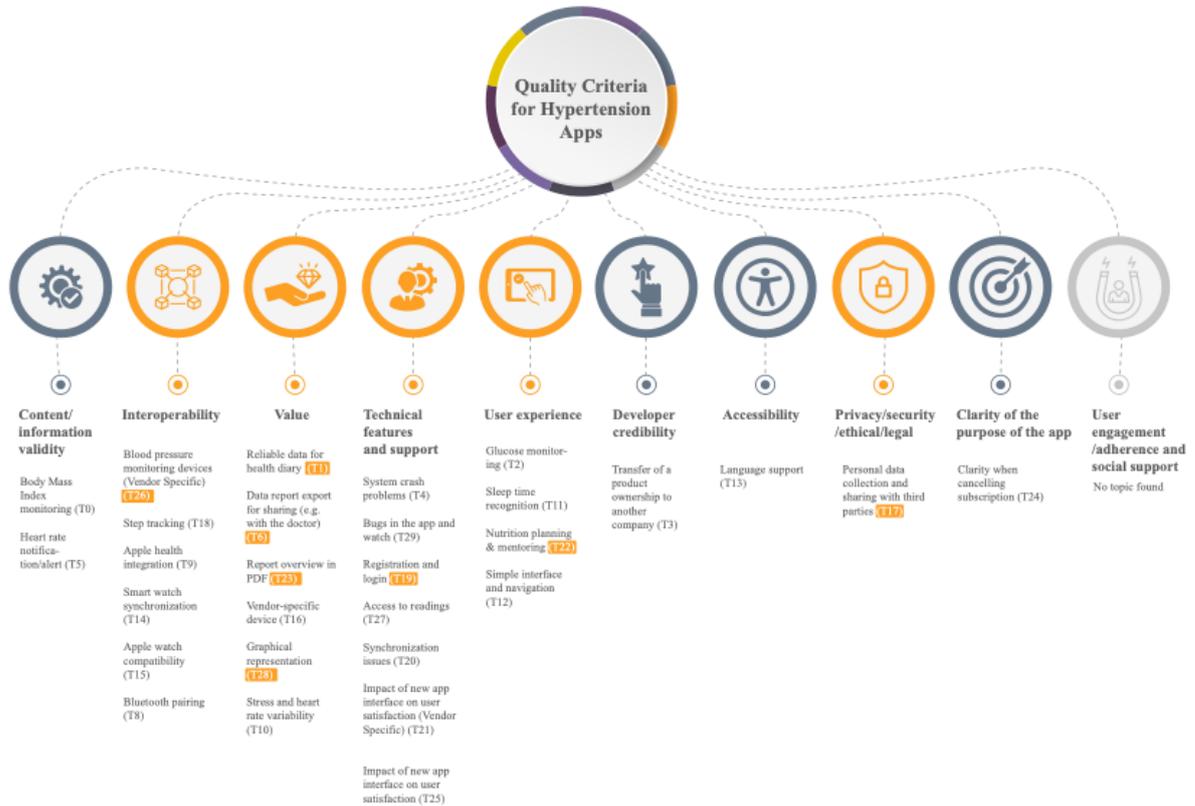

*Figure 5. Topic Modelling Mapped to Evaluation App Framework*

Although the dimension *Content/Information Validity* (1498 reviews) only has two topics, it is the most discussed dimension from our dataset. In particular, only the topic `Body Mass Index Monitoring (BMI)' (T0) has 1078 reviews, being the topic with the most reviews in our dataset. Reviews within this dimension are concerned with accuracy in BMI and heart rate measures (T5). Content/Information Validity refers to readability, credibility, characteristics, quality, and accuracy of the information in the app. The second dimension more discussed was *Interoperability* (1433 reviews). Interoperability refers to the data sharing and data transfer capabilities of the apps. Our dataset indicates that relevant topics for the users include issues related to synchronization when transferring data from blood pressure monitoring devices to the app (T7,T26). Moreover, many reviews seek for a functional integration between the app and smartwatches (T7, T26,T14,T15), and between the app and the Apple Health App (T9). While most topics indicate more general concerns about connectivity issues, one topic refers specifically about issues in tracking when syncing pedometer and the app (T8). The third most frequently discussed dimension was *Value* (N=1391) which refers to perceived benefits and advantages associated with the use of the app. Topics that are discussed in this dimension indicate that users value having an option to export the measurement of health data in PDF (T23) so that it can be shared with health care professionals (T6). Moreover, the export function enables users to have health diary (T1).

*Technical Features and Support* (1058 reviews) refers to defects, errors, quantity and timely updates, and technical support and service quality provided within the app. The reviews within this dimension indicate user concerns about defects in the system, such as the app crashing very often (T4), several bugs in the app and smartwatch (T29), and problems with registration and login to the app (T19). Additionally, users voiced concerns about the impact of the system update on the app in terms of changes in the app interface (T21, T25), and new synchronization issues due to updates (T20). *User Experience* (1057 reviews) refers to the experience of using an app in terms of user-friendliness, design features, functionalities, and ability to provide a personalized experience. Within this dimension, the most mentioned functionalities were glucose monitoring (T2) and sleep time recognition (T22). Additionally, the reviews indicate a need for a personalized experience through mentoring as part of a nutrition program (T22). Preferences for a simple interface and navigation (T12) were also found in the data. *Developer Credibility* (444) refers to the transparency of the app development and testing processes, and accountability and credibility of the app developer, affiliations, and sponsors. Reviews categorized within this dimension indicate concerns over when a company that has a damaged brand image takes over the product of a company with good branding (T3).

Less frequently discussed topics are related to *Accessibility* (183 reviews), *Privacy/Security/Ethical/Legal* (141 reviews), and *Clarity of the Purpose of the* App (104 reviews). Accessibility refers to the ability of the app to capture a wider audience, which is represented in our dataset by availability of language support (T13). Privacy/Security/Ethical/Legal pertains to data protection and legalities of the health app concerning whether the health app adhere to guidelines. Within this dimension, reviews indicate concerns over what type of data is collected from the app and whether the data is shared with third parties (T17). Finally, Clarity of the Purpose of the App is concerned with a clear statement of the intended purpose of the app being provided. Our findings indicate that users seek more clarity concerning subscription plans, in particular to knowing whether they will be charged after canceling a subscription (T24).

Interestingly, we did not find indications for topics related to the dimension of *User Engagement/Adherence* and *Social Support* which refers to the extent of how apps maintain user retention using functionalities such as gamification, forum, behavior technique and social support. It is surprising because using intervention techniques for behavioral change and peer support are fundamental for self-management (Hui et al. 2019; Fallon et al. 2021).

## 4.2 Identifying quality factors that influence user rating

In the last step of our research design, we aimed to answer our RQ2 by investigate what topics are the strongest to influence the user's quality rating. Accordingly, we developed a Random Forest classifier with the 30 topics as input variables, and classified the low and high ratings of the 49 apps into two classes. The accuracy score of the Random Forest classifier was 0.74. Furthermore, the SHapley Additive exPlanation (SHAP) measures were used to capture the influence of each topic on the model output (i.e., user's quality rating), and results can be seen in Figure 6. Each dot on the horizontal axis represents a document of the dataset with blue dots representing low feature value and red dots high features values. The features are ranked from top to bottom; therefore, the first feature has the highest sum of absolute SHAP scores. Feature value represents the likelihood that a document (i.e., a review) belongs to a feature (i.e., a topic). Considering that low feature values represent low likelihood that a document belongs to a feature, the blue dots were not considered in our analysis. Red dots in the right direction represent a feature that influences high rating, whereas red dots in the left direction represent a feature that influences low rating.

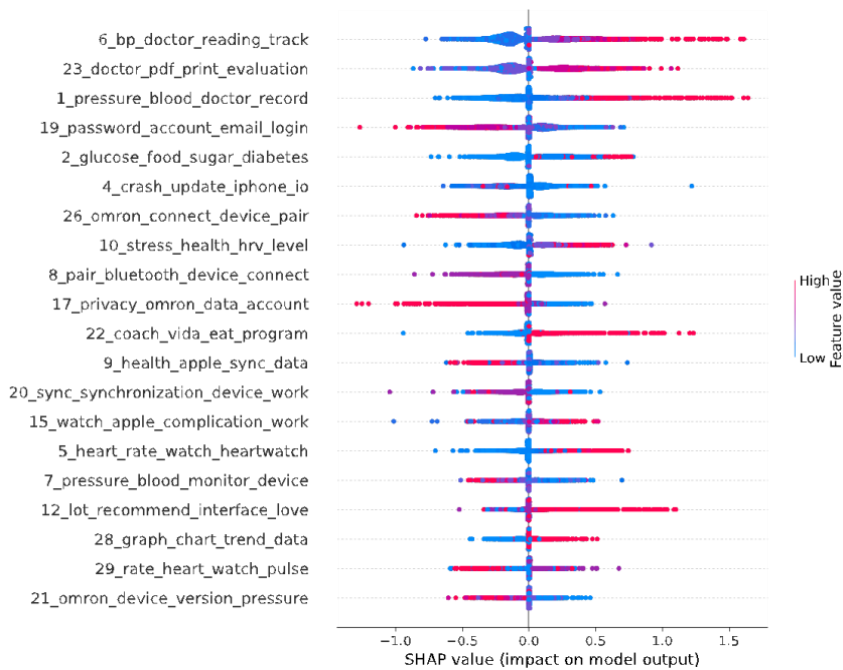

*Figure 6. Feature Importance for the User's Quality Rating*

The graph indicates that the strongest influence on *high* rating are the topics: Report Overview in PDF (T23), Data Report for Sharing (T6), and Reliable Data for Health Diary (T1). Interestingly, the three topics belong to the dimension Value and emphasize the importance of having a feature that enables the user to export health data to not only monitor their condition but also to share the data with their health care professional. Therefore, our dataset indicates that, for hypertension apps, health data export feature is more strongly related to high user's quality rating when compared to the other features. If we turn now to the next topic, having access to mentoring for supporting a healthier diet (T22) has also been reported to strong influence high rating. Although the topic of graphical representation (T28) is not as strong as the three previously mentioned features, it has also been found to influence high rating, and it also enables the user to use health data to monitor and communicate symptoms. The analysis of the reviews indicates a strong user need for mobile apps that enable them to be more empowered over their condition through features that help them to communicate their symptoms to their doctor and that offer access to experts that offers personalized treatment. The five mentioned topics enable the user to have a more in-depth level of education and a deeper level of involvement in the decision-making process related to their condition. We thus propose that:

*Proposition 01: Features that enable effective feedback and communication with health care professionals are the most influential determinants of high-quality rating of hypertension apps.*

The strongest influence on *low* ratings are issues with registration and login (T19), difficulties in synchronizing the app with blood pressure monitoring devices (T26) and privacy concerns related to sharing personal data to third parties (T17). The topics are either related to the dimension of Interoperability, Technical Features and Support, or Privacy. Therefore, our dataset indicates that current hypertension apps are still lacking behind in terms of providing a smooth data exchange between the user's personal devices and a system that is free of defect. As a consequence, important parts of health information of the users are lost which poses risks for them (for example, if a lack of information results in adverse drug interactions). Therefore, ensuring the use of interoperable devices is particularly important in this context. Additionally, manufacturers need to be aware that it is fundamental to be sensitive about the different types of data the app will collect and whether third-party partners will use the data for advertising or analytics. Many reviews indicated a strong need for apps not to collect data that is unnecessary for monitoring the health condition, and neither to share user data with unrelated stakeholders. We thus propose that:

*Proposition 02: The insufficient establishment of privacy concerns and fast health care interoperability resources are the most influential determinants of low-quality rating of hypertension apps.*

## 5 Discussion and Conclusion

This paper contributes to the scholarly literature on mHealth innovation for hypertension. In our study, we operationalized Computational Grounded Theory (Berente & Seidel 2014; Nelson 2020; Ojo & Rizun 2021) to provide insights into quality rating factors that determine the overall user experience from the perspective of the users in the context of mHealth for hypertension. Our findings provide valuable insights and recommendations for the theoretical and practical development of mHealth apps for hypertension.

First, our *findings highlighted the topics that concern the users the most are related to physiological measures, diabetes-related features, and usability aspects*, however; *we found no evidence about psychosocial topics.* For instance, our data clearly indicates that features related to physiological measures such as body mass index monitoring, heart rate variability, stress, and sleep time recognition are relevant for users of hypertension apps. However, it is important to mention that a recent study reported that features related to body monitoring yielded limited efficacy in controlling blood pressure efficacy (K. Liu et al. 2020). Interestingly, features that are normally related to diabetes were also found such as glucose monitoring and nutrition planning & mentoring. This finding is consistent with previous health studies that show that the two diseases are closely interlinked and frequently coexist (K. Liu et al. 2020; Przezak, Bielka & Pawlik 2022). When designing solutions for hypertension, we recommend scholars and practitioners to consider that many of the users might also have diabetes. In contrast to recent studies that promote psychosocial features (e.g. social support) as being important for behavioural change and self-management through mHealth (Hui et al. 2019; K. Liu et al. 2020; Fallon et al. 2021), we found no evidence to support that. This might be explained by the fact that current solutions have mostly focused on improving the accuracy of devices for self-monitoring of hypertension and the display of health data (Kario 2020), while neglecting other components that are relevant for self-management of chronic diseases as explained by (Taylor et al. 2015). Another explanation might be by the fact that scholars have criticized that very few mHealth apps for behavioural change from the Apple App Store are in alignment with behavioural theories (e.g., Self-Determination Theory) (Villalobos-Zúñiga & Cherubini 2020). This creates a clear need for IS scholars to contribute to the discussion and therefore, our study is a call for the community to conduct more research about mHealth for the self-management of hypertension. Our results extends the work of Hui et al. (2019) and Alessa et al. (2018) by providing a list of factors that are relevant for the self-management of hypertension through mHealth from the users' perspective. Finally, many users raised concerns over technical difficulties due to synchronization issues and usability problems. In particular, the lack of a user-friendly design was one of most common mentioned topics. These issues emphasize that the design and development of mHealth apps should follow human-centered design methodologies (Dobrigkeit et al. 2020; Staszak, de Paula & Uebernickel 2021; de Paula, Cormican & Dobrigkeit 2022) to develop solutions that motivates patients to engage in self-care behaviors and further enhance health-related outcomes.

Second, while the mHealth evaluation framework (Hensher et al. 2021). suggests that all criteria are equally important, our model, in contrast, *has predicted that in the context of hypertension, features that enable effective feedback and communication with health care professionals have the strongest influence on the high-quality rating and consequently positive user experience*. Our findings are in alignment with recent developments in health care defined by the new area of digital therapeutics which highlight the need for mHealth to deliver more evidence-based interventions in accordance with health care professionals (Kario 2020). The new era of digital therapeutics brings promises of using more high-quality technologies to develop content that are more personalized, contextualized and just-in-time. So far, IS research is silent about the topic of digital therapeutics with a few exceptions (Lederman et al.

2020; X. Liu et al. 2020). In particular, we recommend scholars to refer to the body of literature on Nudge (Möllenkamp, Zeppernick & Schreyögg 2019) and Just-in-Time Adaptive Interventions (Qian et al. 2022). Furthermore, our *model predicted that the lack of appropriate features for privacy concerns and problematic interoperability functions have the strongest influence on the low-quality rating and consequently negative user experience.* Our work reinforces a previous study that advocate digital medicine to be dependent on interoperability to drive innovation (Lehne et al. 2019). Overall, by predicting that the quality criteria of the mHealth evaluation framework have different importance levels, we extend the work of Horneber & Laumer (2022) and Jakob et al. (2022) by providing a statistical suggestion of factors that influence overall user experience.

Third, our insights provide the basis for inspiring ideas more practical implications for companies include the explanation of a variety of topics that companies could use as basis for inspiring ideas on how to empower users to have a more significant participation in their care and remove obstacles for long-term adherence. It also paves the way for product developers from countries that have committed themselves to reform their health care systems so that all individuals covered by statutory health insurance can be reimbursed for approved digital health applications, such as Germany's new Digital Healthcare Act (Gerke, Stern & Minssen 2020).

Based on our findings, *we make the following three recommendations: (1)* promoting an effective communication between users and health care professionals is associated with high user-rating, and therefore should be prioritized, *(2)* topics that seem to have less influence but still important for the self-management of hypertension are physiological measures (e.g., body mass index monitoring, heart rate variability, stress, and sleep time recognition) and diabetes-related (e.g., glucose monitoring and nutrition planning & mentoring), and *(3)* more focus on providing just-in-time interventions that are aligned with behavioral theories is needed.

This study is not without limitations. The main limitation is concerned with the nature of the methods used which use statistical analysis to suggest predictions. To move towards more formal theories, further validation of and triangulation of results using more traditional quantitative or qualitative approaches is required. Additionally, we that we did not consider potential bias in our dataset. Considering that most reviews are either from Europe or North America and written in English, there might happen that cultural bias influenced our results. Despite these limitations, our paper sets the tone for future studies to design the next generation of mHealth for hypertensions which puts the user in a more active role.

Overall, the propositions for theoretical inquiry and recommendations for practice offered here are mere starting points. Our work makes three important contributions to IS research. Theoretically, our work (a) predicts the relative importance of the quality criteria from the mHealth evaluation framework in the context of hypertension, and (b) it provides empirical evidence for two propositions that can be further developed into more formal theories. Methodologically, we contribute by demonstrating how to operationalize CGT to extract and prioritize knowledge from user reviews from the Apple App Store. Researchers can use our model and open-source tool to investigate other health conditions. Furthermore, it contributes to practice by providing recommendations as a guide for the prioritization of interventions for hypertension apps. Finally, it serves as a call for IS scholars to engage more intensively in scholarly discussions about designing theoretically-driven interventions for hypertension patients.

# References

Alessa, T., Abdi, S., Hawley, M.S. & Witte, L. de, 2018, "Mobile Apps to Support the Self-Management of Hypertension: Systematic Review of Effectiveness, Usability, and User Satisfaction," *JMIR mHealth and uHealth*, 6(7), e10723.

Alhaj, F., Alhaj, A.M., Sharieh, A. & Jabri, R., 2022, "Improving Arabic Cognitive Distortion Classification in Twitter using BERTopic," *International Journal of Advanced Computer Science and Applications*, 13(1).

Alsheref, F.K., 2019, "Medical Information Extraction Model for User-generated Content," *Acta informatica medica: AIM: journal of the Society for Medical Informatics of Bosnia & Herzegovina: casopis Drustva za medicinsku informatiku BiH*, 27(3), 192–198.

Angell, S.Y., De Cock, K.M. & Frieden, T.R., 2015, "A public health approach to global management of hypertension," *The Lancet*, 385(9970), 825–827.

Barlow, J., Wright, C., Sheasby, J., Turner, A. & Hainsworth, J., 2002, "Self-management approaches for people with chronic conditions: a review," *Patient education and counseling*, 48(2), 177–187.

Berente, N. & Seidel, S., 2014, *Big Data & Inductive Theory Development: Towards Computational Grounded Theory?*, Americas Conference on Information Systems (AMCIS).

Bird, S., Klein, E. & Loper, E., 2009, *Natural Language Processing with Python: Analyzing Text with the Natural Language Toolkit*, "O'Reilly Media, Inc."

Bouma, G., 2009, "Normalized (pointwise) mutual information in collocation extraction," *Proceedings of GSCL*.

Breiman, L., 2001, "Random Forests," *Machine learning*, 45(1), 5–32.

Campello, R.J.G.B., Moulavi, D. & Sander, J., 2013, *Density-Based Clustering Based on Hierarchical Density Estimates*, Advances in Knowledge Discovery and Data Mining, 160–172, Springer Berlin Heidelberg.

Cao, W., Milks, M.W., Liu, X., Gregory, M.E., Addison, D., Zhang, P. & Li, L., 2022, "mHealth Interventions for Self-management of Hypertension: Framework and Systematic Review on Engagement, Interactivity, and Tailoring," *JMIR mHealth and uHealth*, 10(3), e29415.

Carter, J.V., Pan, J., Rai, S.N. & Galandiuk, S., 2016, "ROC-ing along: Evaluation and interpretation of receiver operating characteristic curves," *Surgery*, 159(6), 1638–1645.

Choi, J.Y., Choi, H., Seomun, G. & Kim, E.J., 2020, "Mobile-Application-Based Interventions for Patients With Hypertension and Ischemic Heart Disease: A Systematic Review," *The journal of nursing research: JNR*, 28(5), e117.

Debortoli, S., Mueller, O., Junglas, I. & Brocke, J. vom, 2016, "Text Mining For Information Systems Researchers: An Annotated Topic Modeling Tutorial," *Communications of the Association for Information Systems*, 39(7), 110–135.

Devlin, J., Chang, M.-W., Lee, K. & Toutanova, K., 2018, *BERT: Pre-training of Deep Bidirectional Transformers for Language Understanding*, arXiv [cs.CL].

Dieng, A.B., Ruiz, F.J.R. & Blei, D.M., 2020, "Topic modeling in embedding spaces," *Transactions of the Association for Computational Linguistics*, 8, 439–453.

Ding, K., Choo, W.C., Ng, K.Y., Ng, S.I. & Song, P., 2021, "Exploring Sources of Satisfaction and Dissatisfaction in Airbnb Accommodation Using Unsupervised and Supervised Topic Modeling," *Frontiers in psychology*, 12, 659481.

Dobrigkeit, F., Pajak, P., Paula, D. de & Uflacker, M., 2020, "DT@IT Toolbox: Design Thinking Tools to Support Everyday Software Development," in C. Meinel & L. Leifer (eds.), *Design Thinking Research : Investigating Design Team Performance*, pp. 201–227, Springer International Publishing, Cham.

Donkin, L., Christensen, H., Naismith, S.L., Neal, B., Hickie, I.B., Glozier, N. & Others, 2011, "A systematic review of the impact of adherence on the effectiveness of e-therapies," *Journal of medical Internet research*, 13(3), e1772.

Eryilmaz, E. & Thoms B Ahmed, 2022, "Cluster Analysis in Online Learning Communities: A Text Mining Approach," *Communications of the Association for Information Systems,* 51.

Fallon, M., Schmidt, K., Aydinguel, O. & Heinzl, A., 2021, *Feedback Messages During Goal Pursuit: The Dynamic Impact on mHealth Use*, International Conference on Information Systems (ICIS) 2021.

Gerke, S., Stern, A.D. & Minssen, T., 2020, *Germany's digital health reforms in the COVID-19 era: lessons and opportunities for other countries*, npj Digital Medicine, 3(1).

Ghose, A., Guo, X., Li, B. & Dang, Y., 2022, "Empowering patients using smart mobile health platforms: evidence from a randomized field experiment," *MIS Quarterly*, 46(1), 151–191.

Goes, P.B., Lin, M. & Yeung, C.-M.A., 2014, "'Popularity Effect' in User-Generated Content: Evidence from Online Product Reviews," *Information Systems Research*, 25(2), 222–238.

Grootendorst, M., 2022, *BERTopic: Neural topic modeling with a class-based TF-IDF procedure*, arXiv [cs.CL].

Hensher, M., Cooper, P., Wanni, S., Dona, A., Angeles, M.R., Nguyen, D., Heynsbergh, N., Lou Chatterton, M. & Peeters, A., 2021, "Scoping review: Development and assessment of evaluation frameworks of mobile health apps for recommendations to consumers," *Journal of the American Medical Informatics Association: JAMIA*, 28(6), 1318–1329.

Horneber, D. & Laumer, S., 2022, *The app caused me to cancel - Understanding mobile health app dissatisfaction: An affordance perspective dissatisfaction: An affordance perspective*, ECIS.

Hui, C.Y., Creamer, E., Pinnock, H. & McKinstry, B., 2019, "Apps to Support Self-Management for People With Hypertension: Content Analysis," *JMIR mHealth and uHealth*, 7(6), e13257.

Jakob, R., Harperink, S., Rudolf, A.M., Fleisch, E., Haug, S., Mair, J.L., Salamanca-Sanabria, A. & Kowatsch, T., 2022, "Factors Influencing Adherence to mHealth Apps for Prevention or Management of Noncommunicable Diseases: Systematic Review," *Journal of medical Internet research*, 24(5), e35371.

Kario, K., 2020, "Management of Hypertension in the Digital Era: Small Wearable Monitoring Devices for Remote Blood Pressure Monitoring," *Hypertension*, 76(3), 640–650.

Kazuomi, K., Harada, N. & Okura, A., 2022, "Digital Therapeutics in Hypertension: Evidence and Perspectives," *Hypertension*, 79(10).

Lederman, R., D'Alfonso, S., Rice, S., Coghlan, S., Wadley, G. & Alvarez-Jimenez, M., 2020, *Ethical issues in online mental health interventions*, ECIS.

Lehne, M., Sass, J., Essenwanger, A., Schepers, J. & Thun, S., 2019, "Why digital medicine depends on interoperability," *NPJ digital medicine*, 2, 79.

Liu, K., Xie, Z. & Or, C.K., 2020, "Effectiveness of Mobile App-Assisted Self-Care Interventions for Improving Patient Outcomes in Type 2 Diabetes and/or Hypertension: Systematic Review and Meta-Analysis of Randomized Controlled Trials," *JMIR mHealth and uHealth*, 8(8), e15779.

Liu, X., Zhang, B., Susarlia, A. & Padman, R., 2020, "Go to you tube and call me in the morning: Use of social media for chronic conditions," *MIS Quarterly*, 44(1), 257–283.

Lukyanenko, R., Parsons, J., Wiersma, Y., Sieber, R. & Maddah, M., 2016, "Participatory Design for User-generated Content: Understanding the challenges and moving forward," *Scandinavian Journal of Information Systems*, 28(1), 37–70.

Lukyanenko, R., Parsons, J., Wiersma, Y.F. & Maddah, M., 2019, "Expecting the Unexpected: Effects of Data Collection Design Choices on the Quality of Crowdsourced User-Generated Content," *MIS Quarterly*, 43(2), 623–647.

Márquez Contreras, E., Márquez Rivero, S., Rodríguez García, E., López-García-Ramos, L., Carlos Pastoriza Vilas, J., Baldonedo Suárez, A., Gracia Diez, C., Gil Guillén, V. & Martell Claros, N., 2019, "Specific hypertension smartphone application to improve medication adherence in hypertension: a cluster-randomized trial," *Current medical research and opinion*, 35(1), 167–173.

McInnes, L., Healy, J. & Melville, J., 2018, *UMAP: Uniform Manifold Approximation and Projection for Dimension Reduction*, arXiv [stat.ML].

Möllenkamp, M., Zeppernick, M. & Schreyögg, J., 2019, "The effectiveness of nudges in improving the self-management of patients with chronic diseases: A systematic literature review," *Health policy*, 123(12), 1199–1209.

Muntner, P., Hardy, S.T., Fine, L.J., Jaeger, B.C., Wozniak, G., Levitan, E.B. & Colantonio, L.D., 2020, "Trends in Blood Pressure Control Among US Adults With Hypertension, 1999-2000 to 2017-2018," *JAMA: the journal of the American Medical Association*, 324(12), 1190–1200.

Nelson, L.K., 2020, "Computational Grounded Theory: A Methodological Framework," *Sociological methods & research*, 49(1), 3–42.

Ojo, A. & Rizun, N., 2021, *What matters most to patients? On the Core Determinants of Patient Experience from Free Text Feedback Patient Experience from Free Text Feedback*, International Conference on Information Systems (ICIS).

Paula, D. de, Cormican, K. & Dobrigkeit, F., 2022, "From Acquaintances to Partners in Innovation: An Analysis of 20 Years of Design Thinking's Contribution to New Product Development," *IEEE Transactions on Engineering Management*, 69(4), 1664–1677.

Paula, D. de, Zarske Bueno, J. & Viljoen, A., 2021, *Defining Archetypes and Requirements for mHealth Interventions in Rural Kenya: An Investigation in Collaboration with CURAFA$^{TM}$*, Proceedings of the 54th Hawaii International Conference on System Sciences, 4941–4950.

Pedregosa, F., Varoquaux, G. & Gramfort, A., 2011, "Scikit-learn: Machine learning in Python," *Journal of Machine Learning Research*, 12, 2825–2830.

Przezak, A., Bielka, W. & Pawlik, A., 2022, "Hypertension and Type 2 Diabetes—The Novel Treatment Possibilities," *International journal of molecular sciences*, 23(12), 6500.

Qian, T., Walton, A.E., Collins, L.M., Klasnja, P., Lanza, S.T., Nahum-Shani, I., Rabbi, M., Russell, M.A., Walton, M.A., Yoo, H. & Murphy, S.A., 2022, "The microrandomized trial for developing digital interventions: Experimental design and data analysis considerations," *Psychological methods*.

Ruelens, A., 2021, "Analyzing user-generated content using natural language processing: a case study of public satisfaction with healthcare systems," *SIAM journal on scientific computing: a publication of the Society for Industrial and Applied Mathematics*, 1–19.

Savoli, A., Barki, H. & Pare, G., 2020, "Examining how chronically ill patients' reactions to and effective use of information technology can influence how well they self-manage their illness," *MIS Quarterly*, 44(1), 351–389.

Schmitt, F., Sundermeier, J., Bohn, N. & Morassi Sasso, A., 2020, *Spotlight on Women in Tech: Fostering an Inclusive Workforce when Exploring and Exploiting Digital Innovation Potentials when Exploring and Exploiting Digital Innovation Potentials*, ICIS 2020 Proceedings.

Staszak, W., Paula, D. de & Uebernickel, F., 2021, *The power of habits: evaluation of a mobile health solution for the management of narcolepsy*, DS 109: Proceedings of the Design Society: 23rd International Conference on Engineering Design (ICED21), vol. 1, 3081–3090, Cambridge University Press.

Stoyanov, S.R., Hides, L., Kavanagh, D.J. & Wilson, H., 2016, "Development and Validation of the User Version of the Mobile Application Rating Scale (uMARS)," *JMIR mHealth and uHealth*, 4(2), e72.

Taylor, S.J.C., Pinnock, H., Epiphaniou, E., Pearce, G., Parke, H.L., Schwappach, A., Purushotham, N., Jacob, S., Griffiths, C.J., Greenhalgh, T. & Sheikh, A., 2015, *A rapid synthesis of the evidence on interventions supporting self-management for people with long-term conditions: PRISMS – Practical systematic Review of Self-Management Support for long-term conditions*, NIHR Journals Library, Southampton (UK).

Villalobos-Zúñiga, G. & Cherubini, M., 2020, "Apps That Motivate: a Taxonomy of App Features Based on Self-Determination Theory," *International journal of human-computer studies*, 140, 102449.

World Health Organization (WHO), 2021, *Guideline for the pharmacological treatment of hypertension in adults*.

Zhou, B., Carrillo-Larco, R.M., Danaei, G., Riley, L.M., Paciorek, C.J., Stevens, G.A., Gregg, E.W., Bennett, J.E., Solomon, B., Singleton, R.K. & Others, 2021, "Worldwide trends in hypertension prevalence and progress in treatment and control from 1990 to 2019: a pooled analysis of 1201 population-representative studies with 104 million participants," *The Lancet*, 398(10304), 957–980.

Zhou, S., Qiao, Z., Du, Q., Wang, G.A., Fan, W. & Yan, X., 2018, "Measuring Customer Agility from Online Reviews Using Big Data Text Analytics," *Journal of Management Information Systems*, 35(2), 510–539.